\documentclass{ws-procs975x65}            
\usepackage[percent]{overpic}
\usepackage[utf8]{inputenc}
\newcommand{\aref}{\ensuremath{a_{\textrm{ref}}}}
\begin{document}

\begin{flushright}LLNL-JRNL-678485
\end{flushright}

\title{Investigation of the scalar spectrum in SU(3) with eight degenerate flavors}

\author{E.~Rinaldi$^*$ \\ for the Lattice Strong Dynamics (LSD) collaboration}

\address{Nuclear and Chemical Sciences Division, Lawrence Livermore National Laboratory,\\ Livermore, California 94550, United States \\
$^*$E-mail: rinaldi2@llnl.gov}

\begin{abstract}
The Lattice Strong Dynamics collaboration is investigating the properties of a SU(3) gauge theory with $N_f = 8$ light fermions on the lattice.
We measure the masses of the lightest pseudoscalar, scalar and vector states using simulations with the nHYP staggered-fermion action on large volumes and at small fermion masses, reaching $M_{\rho}/M_{\pi} \approx 2.2$.
The axial-vector meson and the nucleon are also studied for the same range of fermion masses.
One of the interesting features of this theory is the dynamical presence of a light flavor-singlet scalar state with $0^{++}$ quantum numbers that is lighter than the vector resonance and has a mass consistent with the one of the pseudoscalar state for the whole fermion mass range explored.
We comment on the existence of such state emerging from our lattice simulations and on the challenges of its analysis.
Moreover we highlight the difficulties in pursuing simulations in the chiral regime of this theory using large volumes.
\end{abstract}

\bodymatter


\section{Introduction}
\label{sec:introduction}

There is increasing evidence that non-abelian gauge theories with a large number of flavors display interesting infrared features that are qualitatively different from QCD.
While experimental results can help our understanding of QCD, it is much harder to study SU(3) gauge theories with many light flavors.
Lattice simulations offer a full non-perturbative description of the low energy regime for these models with controllable and improvable uncertainties.

The Lattice Strong Dynamics (LSD) collaboration is currently performing a high-quality lattice computation of many physical observables that can help our understanding of the infrared dynamics in a SU(3) gauge theory with $N_f=8$ massless fermions.
Such a theory is one of the plausible candidates for a composite Higgs model, based on the walking technicolor scenario\cite{Yamawaki:1985zg}.
Moreover, it is a strongly-coupled system which differs considerably from QCD, and that we have not completely theoretically understood.

With our investigation of the $N_f=8$ theory, we are pursuing two main goals.
On one hand, we want to use numerical lattice simulations to determine the presence of a light flavor-singlet $0^{++}$ scalar state in the massless limit of the theory, since this is a major experimental constraint for the theory to be a viable composite Higgs model.
On the other hand, we want to study this theory \emph{per se} in order to gain insights on the dynamics of strongly-coupled theories near the edge of the conformal window.

So far, there have been a handful of lattice results for the SU(3) gauge theory with $N_f=8$ massless fermions and they will be discussed and referred to in the next sections.
We will show that we are able to improve over existing studies by using lighter fermion masses and larger volumes.
This is a first step towards exploring the limit of massless fermions with controllable systematic and statistical errors.
We remind the reader that all results included in the following come from a preliminary study, presented at the Sakata memorial workshop on strongly-coupled gauge theories (SCGT15) in March 2015.

\section{Lattice setup and scale setting}
\label{sec:lattice-setup-scale}

For our investigation we use the same quark and gauge actions of previous USBSM studies\cite{Schaich:2013eba} focused on the spectrum, and previous works of the Boulder group\cite{Cheng:2011ic}.
The quark action includes two species of degenerate nHYP-smeared staggered quarks with smearing parameter $\alpha=\{0.5,0.5,0.4\}$ and bare mass $m_f$ to describe eight degenerate flavors in the continuum theory.
The SU(3) gauge action includes a fundamental and an adjoint plaquette term with lattice couplings related by $\beta_F = -4 \beta_A$.
The main advantages of this action are the cheaper computational cost with respect to the HISQ action while maintaining a good control over taste-splitting effects, and the possibility of exploring stronger couplings.

The strongest coupling that can be simulated with this action is however limited by the presence of a lattice bulk phase transition.
We choose a lattice gauge coupling $\beta_F=4.8$ that is stronger than the one used in a previous USBSM study\cite{Schaich:2013eba} ($\beta_F=5.0$) but still outside the bulk lattice phase.
We simulate hyper-cubic lattices $L^3\times T$ with $L=24,\ 32,\ 48,\ 64$ and $T=2L$ at small masses $0.00889 \geq m_f \geq 0.00125$.
The numerical simulations of this large-volume and small-mass study are handled by the \emph{QHMC} code that is part of the FUEL project\cite{Osborn:2014kda} and optimized for the BlueGene/Q architecture.

\begin{figure}[ht]
  \begin{tabular}{cc}
    \begin{overpic}[width=0.45\linewidth]{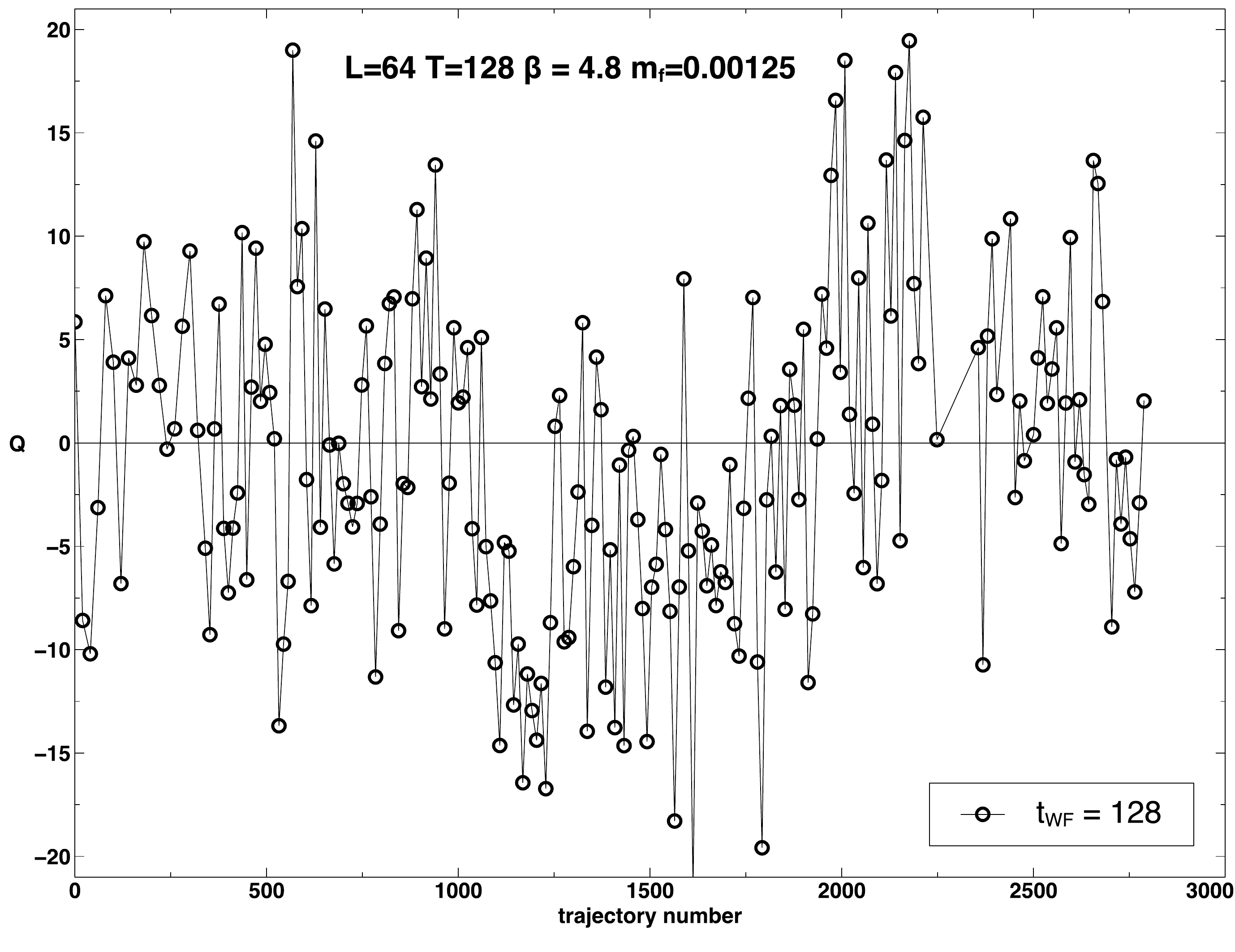}
      \put(10,10){\textbf{PRELIMINARY}}
    \end{overpic} &
                    \begin{overpic}[width=0.45\linewidth]{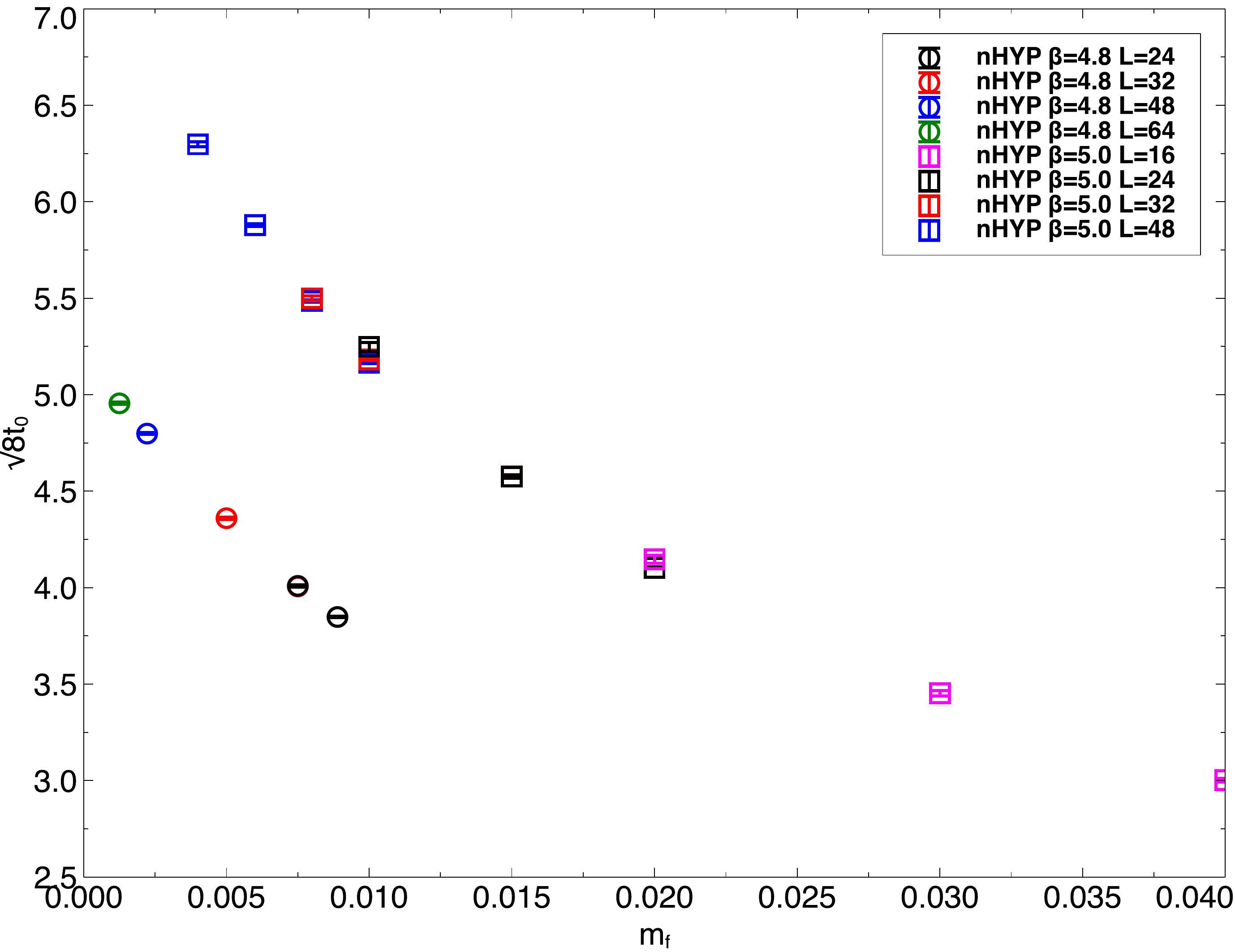} 
                      \put(10,10){\textbf{PRELIMINARY}}
                    \end{overpic} \\
  \end{tabular}
  \caption{\label{fig:topology-and-scale} Left panel: topological charge evolution on the $L=64$ $T=128$ $m_f=0.00125$ ensemble measured at Wilson flow time $t_{\textrm{WF}} = 128$ corresponding to a \emph{smearing} radius of $\sqrt{8t_{\textrm{WF}}} = L/2$. Right panel: lattice scale $t_0$ for our $\beta_F=4.8$ ensembles and for the USBSM ensembles at $\beta_F=5.0$. There is clear evidence of a strong fermion mass dependence, which is absent in simulations with $N_f=2$ or $4$.}
\end{figure}

As a first step we monitor the energy $E$ and the topological charge $Q$ evolution at different Wilson flow times\cite{Luscher:2010iy}.
These gauge observables are used to get a quantitative idea of the quality of the ensembles in terms, for example, of topological tunneling and autocorrelation times.
We observe frequent tunnelling between different topological sectors, even at our lightest quark mass, which is a reassuring sign of ergodicity in the HMC dynamics.
This is shown in the left panel of Fig.~\ref{fig:topology-and-scale}.

We use the energy measurements $E(t)$ along the Wilson flow to define a lattice scale $t_0$ which is the Wilson flow time satisfying the following equation: $t^2\langle E(t) \rangle = c$, where $c=0.3$ is chosen for reference in the following.
The choice of $c$ is arbitrary and it depends on the system under scrutiny.
One interesting observation is that, in this system, the scale $t_0$ is heavily dependent on the fermion mass $m_f$, contrary to the behavior observed at lower number of flavors.

The same lattice scale $t_0$ is defined on the USBSM ensembles at $\beta_F=5.0$ and plotted in the right panel of Fig.~\ref{fig:topology-and-scale}.
By defining a reference length $\aref = \sqrt{8t_0}$ and using dimensionless observables like $\aref M_{\pi}$ we will compare the spectrum from different ensembles.
Moreover we will add results obtained with the HISQ action\cite{Aoki2014} to the comparison.

\section{Exploring the connected spectrum towards the chiral limit}
\label{sec:expl-conn-spectr}

For all fermion masses at $\beta_F=4.8$ we collect more than $\mathcal{O}(10000)$ HMC trajectories, except for $m_f=0.00125$ at $L=64$, where we are only able to generate $\sim 2000$ trajectories due to limited computational resources at this time.
The lattice volumes and masses are chosen with $M_{\pi} L \geq 5.3$ to keep volume effects under control and still be able to reach very light quark masses with available resources.

For the first time we are able to explore, on large volumes, a region of fermion masses where the vector meson ($\rho$) is heavier than twice the pseudoscalar meson ($\pi$), while previous state-of-the-art simulations with the HISQ staggered action\cite{Aoki:2013xza} and more costly domain-wall-fermion simulations\cite{Appelquist2014a} could only achieve $M_{\rho}/M_{\pi} \lesssim 1.5$.

\begin{figure}[ht]
  \begin{tabular}{cc}
    \begin{overpic}[width=0.45\linewidth]{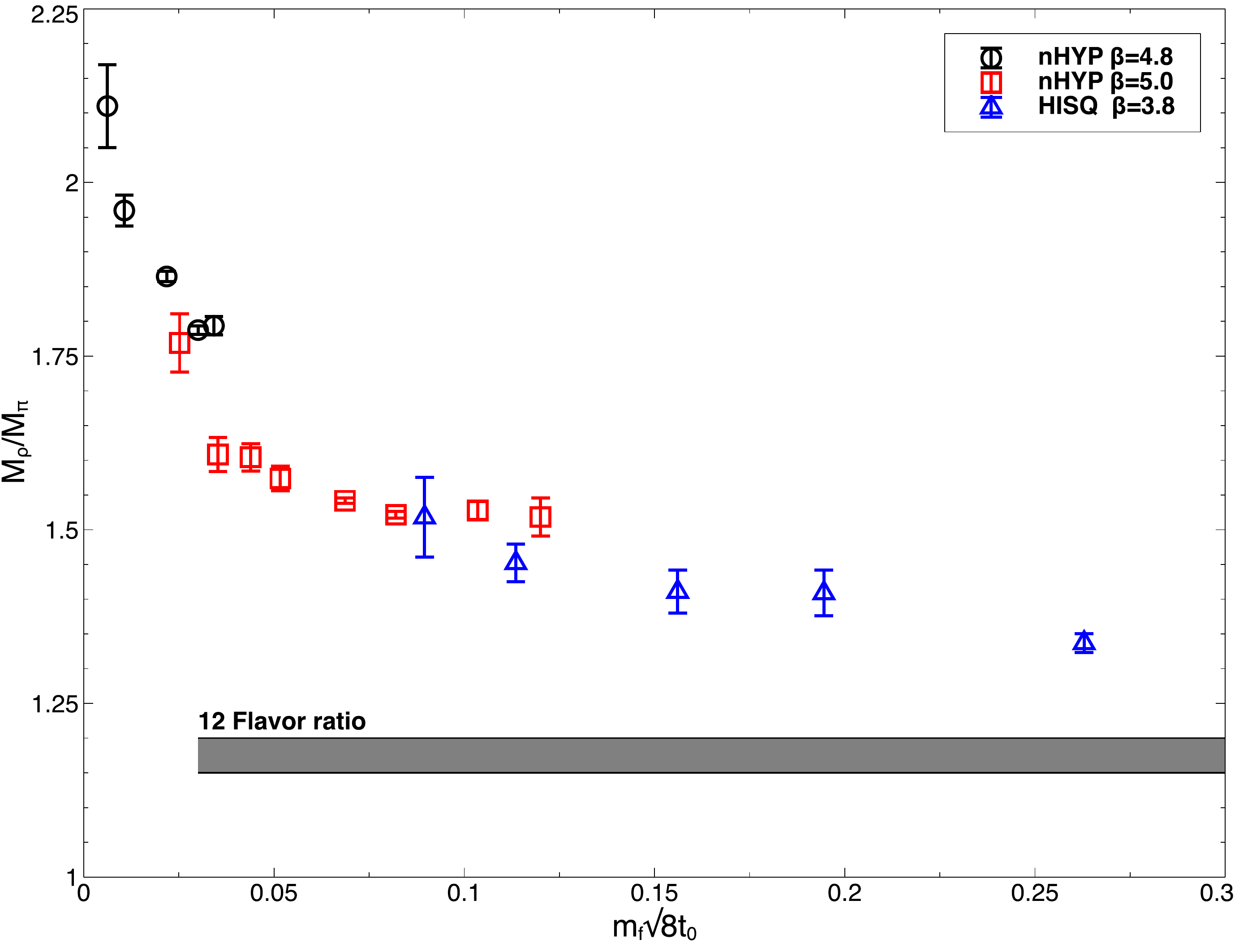}
      \put(15,70){\textbf{PRELIMINARY}}
    \end{overpic} &
                    \begin{overpic}[width=0.45\linewidth]{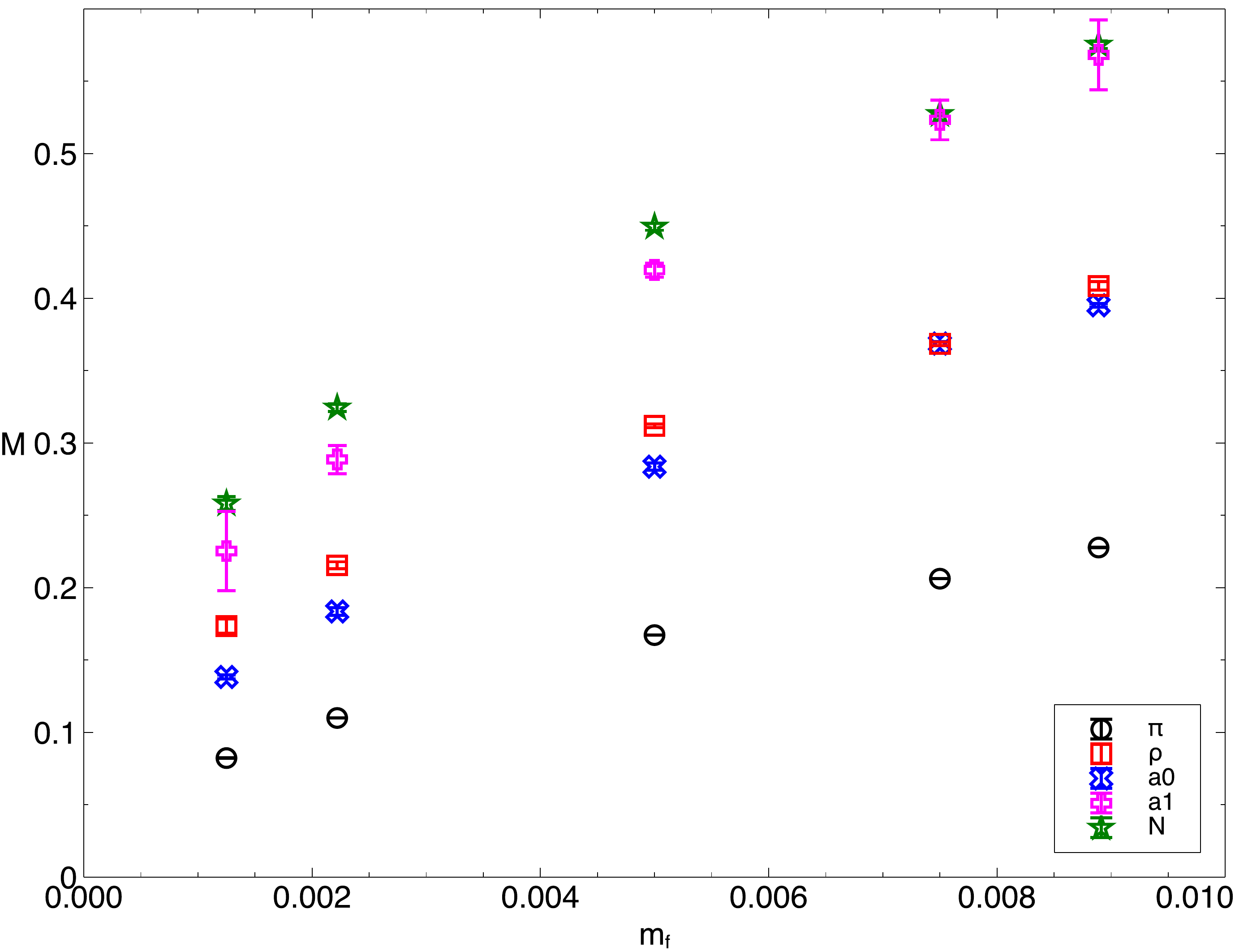}
                      \put(10,70){\textbf{PRELIMINARY}}
                    \end{overpic} \\
  \end{tabular}
  \caption{\label{fig:connected-spectrum} Left panel: The ratio  $M_{\rho}/M_{\pi}$ would diverge in the massless limit of a chirally broken theory. Here we collect this ratio for different gauge couplings and lattice actions, using a rescaled bare fermion mass $\aref m_f$ that allows for a direct comparison. The value of the same ratio which is constant for a $N_f=12$ theory is also shown for comparison. Right panel: The spectrum of connected two-point functions for all the fermion masses and volumes investigated in this study includes the pseudoscalar, vector, scalar, axial-vector and nucleon states.}
\end{figure}

This is a needed and very important step forward in order to understand the long-distance behavior of this gauge theory towards the chiral limit, where $M_{\rho}/M_{\pi} \rightarrow \infty$ if chiral symmetry is spontaneously broken as in QCD.
A clear picture of the progress in this direction is reported in the left panel of Fig.~\ref{fig:connected-spectrum}, where we collect our results at $\beta_F=4.8$, together with the ones from the nHYP study at $\beta_F=5.0$\cite{Schaich:2013eba} and from a HISQ study\cite{Aoki:2013xza}.
We are able to qualitatively compare on the same plot different coupling constants for the nHYP action, as well as different lattice discretizations, thanks to a common reference scale $\aref$ (see previous section) which we use to rescale the bare fermion mass $m_f$.
The same plot includes the value\cite{Cheng2014}  of $M_{\rho}/M_{\pi}$ for a SU(3) gauge theory with $N_f=12$ which is known to have an infrared-conformal fixed point\cite{Cheng2014a} and displays hyperscaling when it is mass-deformed such that all mass ratios are constant in the massless limit.

The right panel of Fig.~\ref{fig:connected-spectrum} shows a summary of the connected spectrum on all our ensembles.
The lightest state is the pseudoscalar one, followed by the non-singlet scalar ($a_0$) which becomes degenerate with the vector at large fermion masses.
Heavier states include the axial-vector meson ($a_1$) and the nucleon ($N$).

\section{The flavor-singlet scalar channel}
\label{sec:flav-singl-scal}

One of the most interesting spectroscopic observables in the SU(3) gauge theory with $N_f=8$ flavors is the flavor-singlet scalar channel, where a particle with the same quantum numbers of the Higgs boson ($0^{++}$) has been found to be as light as the pseudoscalar state\cite{Aoki2014}.
Such a $0^{++}$ state happens to be dynamically light in other gauge theories with a variety of different number of flavors\cite{Aoki:2013zsa,Brower:2014ita} and even fermion representations\cite{Fodor:2014pqa,DelDebbio:2010hx,Athenodorou:2014eua} for SU(3) and SU(2): the common feature among these theories is that they are all close or inside the conformal window, where their beta function would have a non-trivial infrared conformal fixed point.

Studying the $0^{++}$ channel requires extra care due to the non-zero overlap with the gauge vacuum fluctuations and the presence of disconnected contributions.
From previous QCD studies, it is known that a larger amount of gauge configurations is needed compared to the connected correlation functions.
Moreover, specific techniques to reduce the computational cost associated with disconnected diagrams have to be employed.
There has been a great deal of development in computing all-to-all fermionic propagators needed for this type of calculations and in particular we use U(1) full-volume stochastic sources with full dilution\cite{Foley:2005ac} in time and color, and with even-odd dilution in space.
The scalar interpolating operator $\mathcal{O}_S(t) = \bar{\psi}\psi(t)$ used in our measurements and the technical details of constructing its correlators are the same used for a companion project\cite{Brower:2014ita}.

\begin{figure}[ht]
  \center
  \begin{overpic}[width=0.50\linewidth]{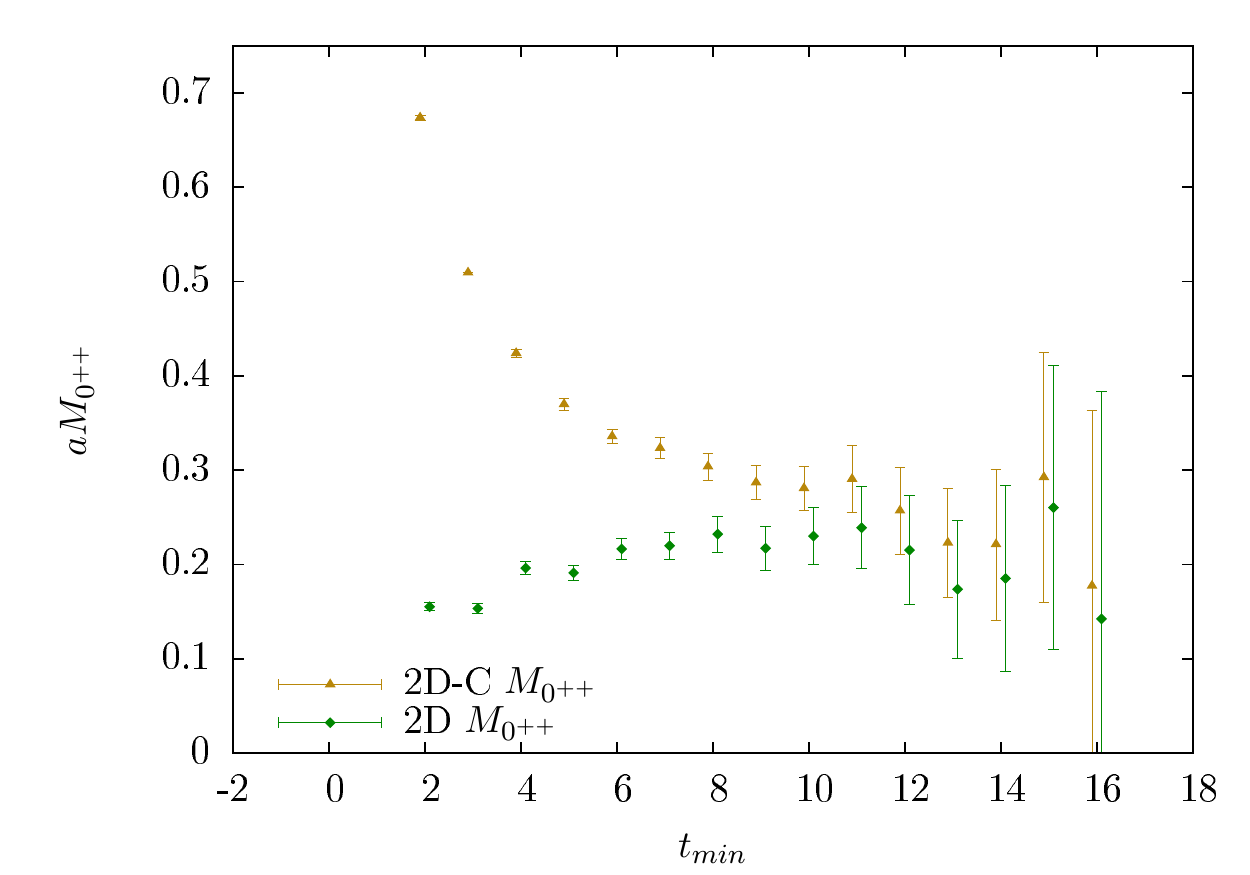}
    \put(45,55){\textbf{PRELIMINARY}}
  \end{overpic}
  \caption{\label{fig:comparison} A fit to the $2D(t)-C(t)$ and $2D(t)$ correlators for $L=24$ and $m_f=0.00889$ using the function in Eq.~(\ref{eq:fit}) and for different fitting windows $[t_{\textrm{min}},T/2]$. At large $t_{\textrm{min}}$ both fits give a consistent answer, albeit with large statistical errors.}
\end{figure}
Let us define two primary correlation functions, $C(t)$ and $D(t)$, which represent the fully-connected correlator and the vacuum-subtracted disconnected correlator, respectively.
While the connected $C(t)$ correlator contains contributions from non-singlet states, the disconnected $D(t)$ one contains also flavor-singlet contributions.
Due to the staggered nature of the fermionic operators, both scalar correlation functions couple to the respective pseudoscalar parity-partner as well.
However, if we call $\sigma$ the lightest state in the scalar channel, the following two combinations will feature an asymptotic large-time behavior dominated by $M_\sigma$:
\begin{eqnarray}
  \label{eq:correlators}
  S(t) \; \equiv \; 2D(t) - C(t) & \approx & c_{\sigma} e^{-M_{\sigma}t} \qquad \textrm{when} \; t \rightarrow \infty \\
  2D(t) & \approx & c_{\sigma} e^{-M_{\sigma}t} \qquad \textrm{when} \; t \rightarrow \infty \ ,
\end{eqnarray}
where the factor of two in front of $D(t)$ represents the number of staggered species.\\
We fit both $S(t)$ and $2D(t)$ with the following fit function
\begin{equation}
  \label{eq:fit}
  F(t) \; = \; c_{0^{++}} \cosh{M_{0^{++}} (t - T/2)} + (-1)^t \ c_1 \cosh{M_1 (t - T/2)} + v_0 \ ,
\end{equation}
that includes a primary state with mass $M_{0^{++}}$, a parity partner state with mass $M_1$.
We also include a free constant parameter $v_0$ representing our inability to precisely resolve the vacuum contribution in $D(t)$ with our limited statistics, in particular at large temporal distance.
The plot in Fig.~\ref{fig:comparison} shows how $M_{0^{++}}$ depends on the window of time-slices $[t_{\textrm{min}},T/2]$ considered in the fit for both correlators.
When contributions from excited states at low $t_{\textrm{min}}$ are removed, we note that $S(t)$ and $2D(t)$ contain a propagating ground state with the same mass: this mass is identified as $M_{\sigma}$ and is the mass of the lightest flavor-singlet $0^{++}$ state.
This comparison is reassuring and it provides, \emph{a posteriori}, a justification for using, in the following, the results coming from $2D(t)$ given that this correlation function appears to be less contaminated by excited states and with a smaller statistical error.

\begin{figure}[ht]
  \begin{tabular}{cc}
    \begin{overpic}[width=0.45\linewidth]{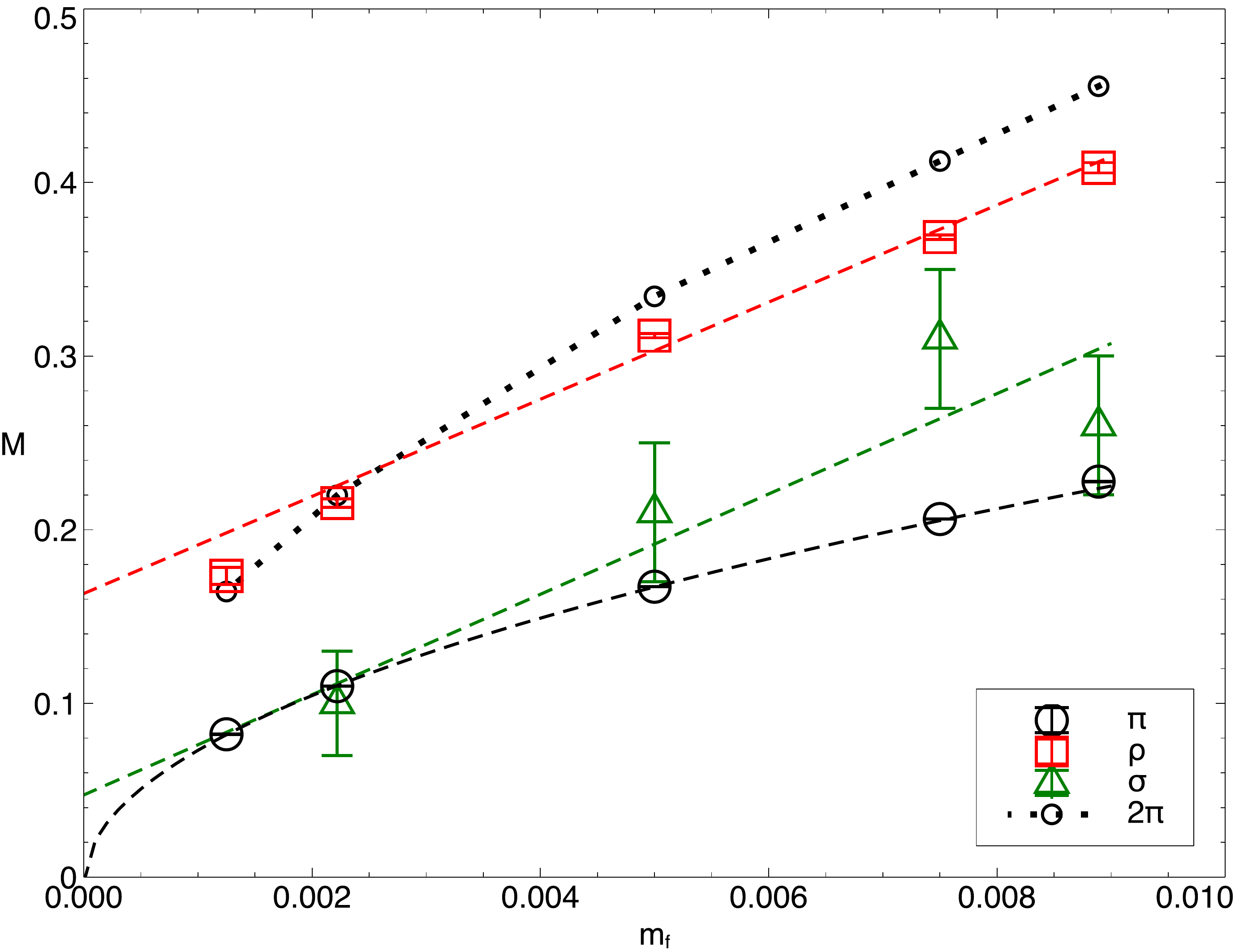}
      \put(10,70){\textbf{PRELIMINARY}}
    \end{overpic} &
                    \begin{overpic}[width=0.45\linewidth]{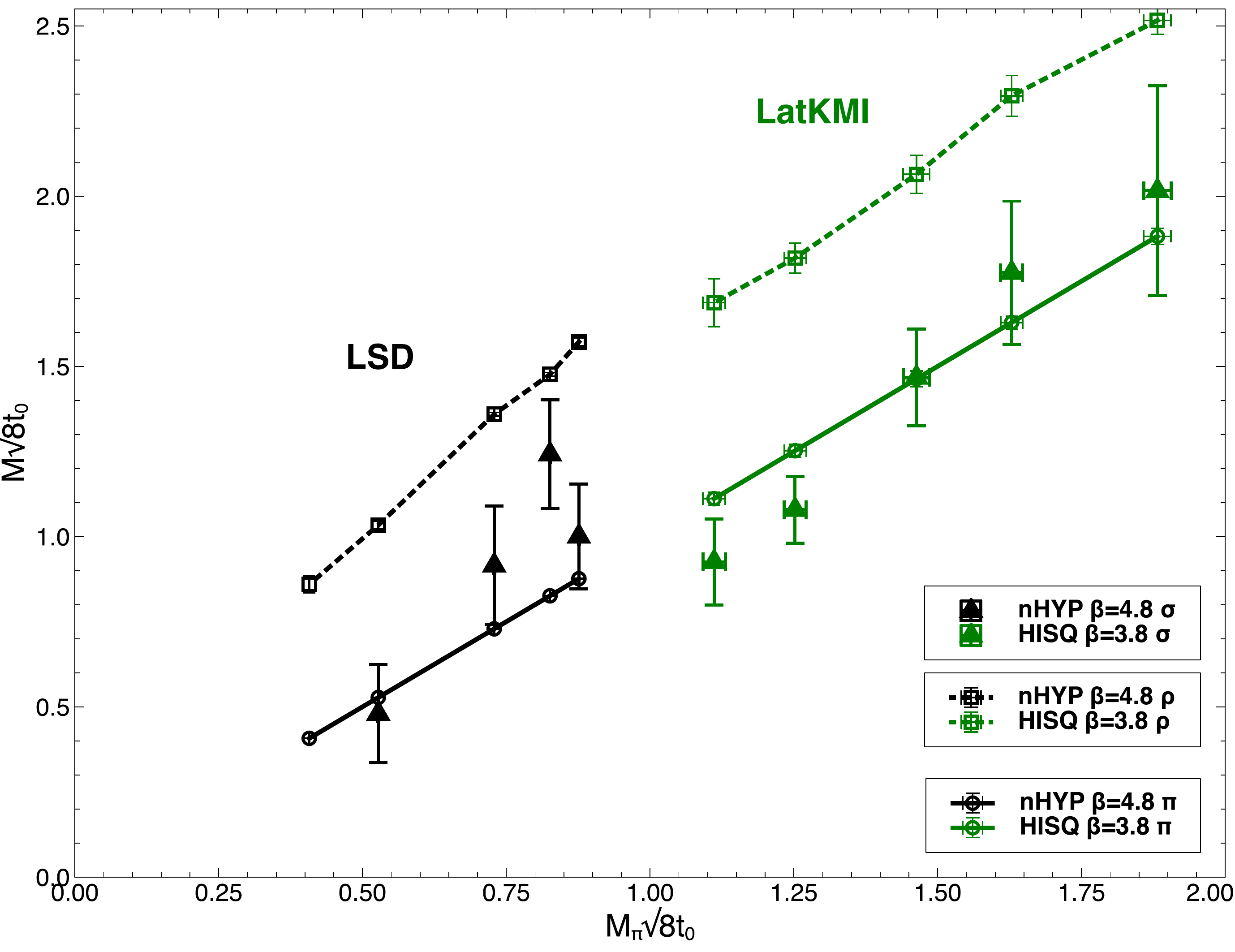} 
                      \put(10,70){\textbf{PRELIMINARY}}
                    \end{overpic} \\
  \end{tabular}
  \caption{\label{fig:scalar-channel} Left panel: The low-lying spectrum of the eight-flavor theory, including the pseudoscalar $\pi$, the vector $\rho$ and the $0^{++}$ flavor-singlet scalar $\sigma$ state. The dotted line represents the $2\pi$ threshold. Dashed lines are for illustration purpose only and guide the eye towards the massless limit of the theory.  Right panel: Comparison, in $\aref$ units, between the $\pi$, $\rho$ and $\sigma$ masses in this work with nHYP fermions (LSD) and in the work with HISQ fermions\cite{Aoki2014} (LatKMI). The pseudoscalar mass is used as an indication of the distance from the massless limit. Solid lines are $\pi$, dotted lines are $\rho$ and triangles are $\sigma$.}
\end{figure}

In Fig.~\ref{fig:scalar-channel} we show a summary of $M_{\pi}$, $M_{\sigma}$ and $M_{\rho}$ on our ensembles.
In the left panel, these masses are compared to the $2M_\pi$ threshold: for all masses but the lightest two, the $\rho$ is below threshold, while $\sigma$ is always below threshold.
Moreover, at light fermion masses, $M_{\sigma}$ is compatible with $M_{\pi}$, even though statistical errors are large.
For the lightest $m_f$, calculations have not yet been performed.
This particular result confirms previous observations by the LatKMI collaboration\cite{Aoki2014} and a comparison is shown in the right panel of Fig.~\ref{fig:scalar-channel}, using a common reference scale.

It is clear that this work explores a regime with considerably lighter fermions.
Nonetheless a light flavor-singlet $0^{++}$ scalar continues to be a prominent feature of the spectrum.
Whether this feature survives in the massless limit is one of the major questions we are addressing using the large computational resources at our disposal.
However, going to lighter fermion masses requires a significant increase in CPU time and might not be feasible.

\section{Conclusions}
\label{sec:conclusions}

With the goal to shed light on the $N_f=8$ SU(3) theory as a phenomenologically viable composite Higgs model and to understand its low-energy dynamics, the LSD collaboration is pursuing large scale simulations of the zero-temperature hadron spectrum with emphasis on the flavor-singlet scalar particle. 
While the results presented in the previous sections are still preliminary, it is outstanding that a unified picture is beginning to emerge by combining efforts from different collaborations.

There are still unanswered questions ranging from the behavior of the non-perturbative beta function with massless quark, to the fate of the $0^{++}$ state in the chiral limit.
Moreover, the existence of a finite-temperature phase transition is still an open question\cite{Schaich:2015psa}.
Our future plans, beside completing the analyses presented here, include a careful comparison of the scalar spectrum and of the beta function with $N_f=4$, which is our template for QCD.
Moreover, we want to expand our previous study\cite{Appelquist2014a} of the $S$ parameter to investigate the electroweak constraints on this theory in the massless limit.

\section*{Acknowledgments}
\label{sec:acknowledgments}
We are grateful to the LatKMI collaboration for sharing their unpublished results on the Wilson flow scale $t_0$.
We thank Anna Hasenfratz and all the members of the LSD collaboration who contributed to the studies discussed in this proceeding.
We acknowledge the Lawrence Livermore National Laboratory (LLNL) Multiprogrammatic and Institutional Computing program for Grand Challenge allocations on the LLNL BlueGene/Q supercomputer.
Additional numerical analyses were carried out on clusters at LLNL, Boston University and DOE-funded USQCD resources at Fermilab.
ER was supported by LLNL LDRD~13-ERD-023 ``Illuminating the Dark Universe with PetaFlops Supercomputing'' and acknowledges the support of DOE under Contract DE-AC52-07NA27344 (LLNL).

\bibliographystyle{ws-procs975x65}
\bibliography{scgt15_rinaldi}

\end{document}